\documentclass[usenatbib]{emulateapj}
\usepackage{graphicx}
\usepackage{natbib}
\usepackage{times}
\usepackage{amssymb}

\defcitealias{ls08}{LS08}
\defcitealias{pf10}{PF10}
\defcitealias{bfp10a}{BFP}

\shorttitle{$\sigma$-$\mathcal{M}$ relation in supersonic turbulence}
\shortauthors{Price, Federrath \& Brunt}

\begin{document}
\label{firstpage}

\title{The density variance -- Mach number relation in supersonic, isothermal turbulence}

\author{Daniel J. Price}
\affil{Centre for Stellar and Planetary Astrophysics, School of Mathematical Sciences, Monash University, Clayton Vic 3168, Australia}
\email{daniel.price@monash.edu}

\author{Christoph Federrath\altaffilmark{1}}
\affil{Zentrum f\"ur Astronomie der Universit\"at Heidelberg, Institut f\"ur Theoretische Astrophysik, Albert-Ueberle-Str.~2, D-69120 Heidelberg, Germany}


\author{Christopher M. Brunt}
\affil{School of Physics, University of Exeter, Stocker Rd, Exeter EX4 4QL, UK}

\altaffiltext{1}{Ecole Normale Sup\'{e}rieure de Lyon, CRAL, 69364 Lyon Cedex 07, France}

\begin{abstract}
We examine the relation between the density variance and the mean-square Mach number in supersonic, isothermal turbulence, assumed in several recent analytic models of the star formation process. From a series of calculations of supersonic, hydrodynamic turbulence driven using purely solenoidal Fourier modes, we find that the `standard' relationship between the variance in the log of density and the Mach number squared, i.e., $\sigma_{\ln \rho/\bar{\rho}}^{2}=\ln\left(1+b^{2}\mathcal{M}^{2}\right)$, with $b = 1/3$ is a good fit to the numerical results in the supersonic regime up to at least Mach~20, similar to previous determinations at lower Mach numbers. While direct measurements of the variance in linear density are found to be severely underestimated by finite resolution effects, it is possible to infer the linear density variance via the assumption of log-normality in the Probability Distribution Function. The inferred relationship with Mach number, consistent with $\sigma_{\rho/\bar{\rho}}\approx b\mathcal{M}$ with $b=1/3$, is, however, significantly shallower than observational determinations of the relationship in the Taurus Molecular Cloud and IC5146 (both consistent with $b\approx 0.5$), implying that additional physics such as gravity is important in these clouds and/or that turbulent driving in the ISM contains a significant compressive component. Magnetic fields are not found to change this picture significantly, in general reducing the measured variances and thus worsening the discrepancy with observations.
\end{abstract}

\keywords{turbulence ---  ISM: structure --- hydrodynamics --- stars: formation --- magnetohydrodynamics (MHD) --- shock waves}

\section{Introduction}
\label{sec:intro}
 The last few years have seen an increasing number of analytic models of the star formation process that use the log-normal density probability distribution function (PDF) produced by supersonic turbulent flows to predict statistical quantities such as the initial and/or core mass function \citep[e.g.][]{pn02,hc08,hennebellechabrier09} and the star formation rate \citep{km05,pn09}. A key assumption in these models is a relationship identified in early numerical studies between the PDF width -- the density variance or standard deviation -- and the Root Mean Square (RMS) Mach number $\mathcal{M}$ in supersonic, isothermal turbulence.  The relationship is generally assumed to be linear in the standard deviation of linear density, i.e.,
\begin{equation}
\sigma_{\rho/\bar{\rho}} =b\mathcal{M},
\label{eq:bfaclin}
\end{equation}
where $b$ is a constant of order unity and density is scaled in terms of the mean, $\bar{\rho}$. For a log-normal distribution, this is equivalent to
\begin{equation}
\sigma_{s}^{2}=\ln\left(1+b^{2}\mathcal{M}^{2}\right),
\label{eq:bfac}
\end{equation}
where $s\equiv\ln(\rho/\bar{\rho})$, such that $\sigma_{s}$ is the standard deviation in the \emph{logarithm of density}.

 Apart from the early empirical findings of \citet{vs94}, \citet*{pnj97} and \citet*{pvs98}, there is no clear reason why the relationship should be of this form. Mathematically, the appearance of a log-normal distribution can be understood as a consequence of the multiplicative central limit theorem assuming that individual density perturbations are independent and random \citep{vs94,pvs98,np99}. In physical terms this has been interpreted as meaning that density fluctuations at a given location are constructed by successive passages of shocks with a jump amplitude independent of the local density \citep[e.g.][]{bpetal07,kritsuketal07,federrathetal10}. However it has not so far proved possible to analytically predict the relationship based on these ideas (though see \citealt{pn09}). Thus, a common approach in numerical studies of turbulence --- usually at a fixed Mach number --- has been to measure the parameter $b$, assuming Eq.~(\ref{eq:bfac}), that gives best fitting log-normal to the time averaged PDF. However, reported estimates for $b$ are widely discrepant. For example, \citet{pnj97} found $b\approx0.5$ while more recently \citet{kritsuketal07} (at Mach 6) find a much lower value of $b\approx 0.26$ and \citet{beetzetal08} find $b\approx0.37$, while \citet{pvs98} found $b\approx1$ (though with some confusion over $\sigma_{s}$ vs. $\sigma_{\rho/\bar{\rho}}$).

  \citet*{fks08} and \citet{federrathetal10} reconcile these results in part by the finding that the width of the PDF depends not only on the RMS Mach number but also on the relative degree of compressible and solenoidal modes in the turbulence forcing, with $b=1/3$ appropriate for purely solenoidal and $b=1$ for purely compressive forcing. This is in keeping with earlier discussions by \citet{pvs98} and \citet{np99}, the latter authors noting that ``for compressional forcing at low Mach numbers (leading to an ensemble of sound waves), the standard deviation is expected to be equal to the RMS Mach number itself''.

 Observationally, the log-normality of the 3D PDF is reflected in the 2D column density PDF, for example as measured from dust extinction maps \citep[e.g.][]{lombardietal06,lombardietal08,lombardietal10,kainulainenetal09} -- at least in earlier stages of molecular cloud evolution, suggesting that this phase could be dominated by roughly isothermal turbulence in which self-gravity is relatively unimportant. Only for seemingly more evolved clouds (including Taurus) do \citet{kainulainenetal09} see significant tails at higher (column) densities (similarly found by \citealt{lombardietal10}). However, measurements of the projected 2D variance (or PDF) cannot be directly used to constrain the relationship with Mach number. Recently, \citet*{bfp10a} (hereafter \citetalias{bfp10a}) have shown how projection effects can be overcome to infer the 3D density variance from column density observations, in turn leading to a method for extracting the unprojected (3D) density PDF from the observational data \citep*{bfp10b}. This enables the relationship between the standard deviation in linear density and Mach number to be tested observationally, with initial application to Taurus finding $b=0.48^{+0.15}_{-0.11}$ \citep{brunt10}. A similar method was employed by \citet*{pjn97} to infer the 3D density variance from extinction measurements in IC5146, similarly finding $b\approx 0.5$.

  The problem with all of the above is that calculations --- or observations --- performed at a single (RMS) Mach number can only ever \emph{assume} the relationship given by Eqs.~(\ref{eq:bfac}) or (\ref{eq:bfaclin}) and cannot be used to constrain it unless a range of Mach numbers are studied. Indeed, \citet{ls08} (hereafter \citetalias{ls08}) --- performing a series of calculations with Mach numbers in the range $1.2\le\mathcal{M}\le6.8$ --- find a relationship
\begin{equation}
\sigma^{2}_{s}=0.72\ln\left(1+0.5\mathcal{M}^{2}\right)-0.20,
\label{eq:ls08}
\end{equation}
 based on a fit to measurements of the mean in the logarithm of density, $\bar{s}$, as a function of $\mathcal{M}$, which we have here converted to a $\sigma_{s}$--$\mathcal{M}$ relation using $\bar{s}=-\sigma_{s}^{2}/2$. However, this three-parameter fit is clearly not unique, and it remains to be determined whether this or a similar relationship continues to hold at higher Mach numbers.
 
  The present study is motivated by a need to compare the theoretical predictions with the observational constraints. In particular \citetalias{ls08} only perform calculations up to $\mathcal{M}\approx6.8$ -- corresponding to 1D line Full-Width-Half-Maximum of $\sim1.6$ km/s (at 10K), which is rather low in terms of what is found in the real interstellar medium. In particular Taurus has $\mathcal{M}\sim17$,  so a study going up to (at least) Mach~20 or so is needed. Our aim in this paper is precisely this: To pin down the theoretical relationship -- with as few assumptions as possible -- up to sufficiently high Mach numbers that a meaningful comparison can be made with observed molecular clouds. Whilst additional physics such as non-isothermality \citep[e.g.][]{scaloetal98}, the multiphase nature of the interstellar medium and self-gravity \citep[e.g.][]{klessen00,kritsuketal10} are all expected to change the theoretical predictions at some level, the isothermal, non-self-gravitating case is an important reference point that remains theoretically uncertain. Furthermore, a clear prediction for this simple case can be used to gauge the relative importance of such additional physics in observed clouds.

\section{Methods}

\subsection{Log-normal distributions}
The log-normal distribution is given by
 \begin{equation}
p(s)ds=\frac{1}{\sqrt{2\pi\sigma_{s}^{2}}}\exp{\left[-\frac12\left(\frac{s-\bar{s}}{\sigma_{s}}\right)^{2}\right]}ds,
\label{eq:lognormal}
\end{equation}
where $s\equiv\ln(\rho/\bar{\rho})$ such that $\bar{s}$ and $\sigma_{s}$ denote the mean and standard deviation in the logarithm of (scaled) density, respectively, and $\bar{\rho}$ is the mean in the linear density. The mean and variances in a log-normal distribution are related by
\begin{equation}
\bar{s}=-\frac12\sigma_{s}^{2},
\label{eq:mus}
\end{equation}
and
\begin{equation}
\sigma_{\rho/\bar{\rho}}^{2}= \exp\left(\sigma^{2}_{s}\right)-1.
\label{eq:sigmas}
\end{equation}

\subsection{Numerical simulations}
We have performed a series of calculations of supersonic turbulence, solving the equations of compressible hydrodynamics 
%
%
using an isothermal equation of state (with sound speed $c_{s}=1$) and periodic boundary conditions in the three-dimensional domain $x,y,z\in[0,1]$. Initial conditions were a uniform density medium $\rho=\bar{\rho}=1$ with zero initial velocities. Turbulence was produced by adding a random, correlated stirring force, driving the few largest Fourier modes $1\!<\!k\!<\!3$ with a random forcing pattern, slowly changed according to an Ornstein-Uhlenbeck (OU) process, such that the pattern evolves smoothly in space and time \citep{schmidt09,federrathetal10}. The driving, and the \textsc{phantom} Smoothed Particle Hydrodynamics (SPH) code employed, are described in detail in \citet{pf10} (PF10, see also \citealt{federrathetal10}). Calculations were evolved for 10 dynamical times [defined as $t_{d}\equiv L/(2\mathcal{M}c_s)$], using only results after $2t_{d}$ such that turbulence is fully established \citep[e.g.][]{fks09}.
The amplitude of the driving force was adjusted to give RMS Mach numbers in the range $1\le\mathcal{M}\le20$ by varying the energy input per Fourier mode proportional to the Mach number squared, i.e., $E_{stir}\propto\mathcal{M}^{2}$ while the correlation time for the OU process was set to $t_{d}$ (for the nominally input $\mathcal{M}$).

 Most importantly, unless otherwise specified we have driven the turbulence using purely solenoidal Fourier modes. Thus, according to the heuristic theory of \citet{federrathetal10} we should expect a relationship of the standard form (\ref{eq:bfac}) with $b\approx1/3$.

\begin{figure}
\begin{center}
 \includegraphics[width=\columnwidth]{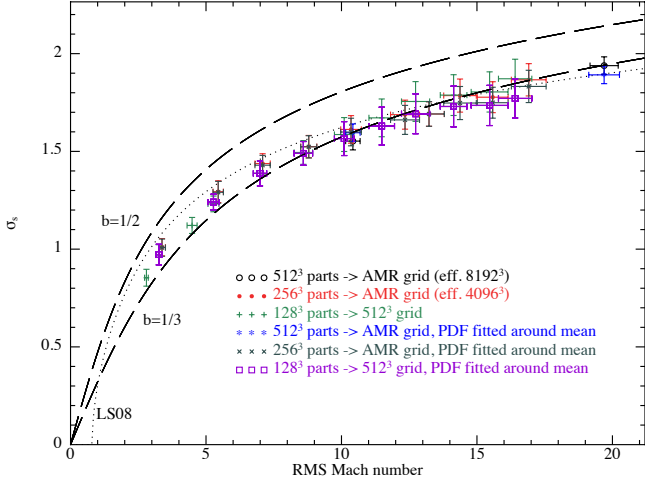}
 \caption{Measured relationship between the (volume-weighted) standard deviation of the logarithm of density $\sigma_{s}$ as a function of RMS Mach number from a series of solenoidally-driven supersonic turbulence calculations. The points show time averages, with error bars showing (temporal) $1\sigma$ deviations. The dashed lines show the standard relation (Eq.~\ref{eq:bfac}) with $b=1/3$ and $b=1/2$ while the dotted line shows the best fitting relationship found by \citet{ls08} (Eq.~\ref{eq:ls08}). Differences between directly measuring $\sigma_{s}$ (open circles, filled circles, and plus signs) compared to fitting the PDF around the mean ($*$,$\times$ and squares) are not significant (i.e., smaller than the time-dependent fluctuations). Overall, the results are consistent with $b=1/3$, as expected for solenoidally-driven turbulence from \citet{fks08,federrathetal10}, and indistinguishable from the \citetalias{ls08} best fit.}
\label{fig:sigmamach}
\end{center}
\end{figure}

\subsection{Measuring the density variance}
We consider a range of methods for measuring the density variance from the simulations.
\begin{enumerate}
\item[i)] Measure the linear variance, $\sigma_{\rho/\bar{\rho}}^{2}$, directly -- with no assumptions about log-normality or otherwise -- and fit the measured relation as a function of $\mathcal{M}$.
\item[ii)] Measure the logarithmic variance $\sigma_{s}^{2}$ directly and fit the measured relation. Infer $\sigma_{\rho/\bar{\rho}}$ assuming a log-normal PDF via Eq.~(\ref{eq:sigmas}).
\item[iii)] Measure $\bar{s}$ and fit the measured relation. Infer $\sigma_{s}$ using Eq.~(\ref{eq:mus}) and in turn $\sigma_{\rho/\bar{\rho}}$ using Eq.~(\ref{eq:sigmas}).
\item[iv)] Determine the value of $\sigma_{s}$ that gives the best fitting PDF in a restricted range around the mean. Infer $\sigma_{\rho/\bar{\rho}}$ using Eq.~(\ref{eq:sigmas}).
\end{enumerate}
The objection to method i) is that it is sensitive to the tails of the density distribution, where time-dependent fluctuations and intermittency effects can cause deviations from log-normality (\citealt{kritsuketal07,federrathetal10}, \citetalias{pf10}). On the other hand no assumptions are made regarding the PDF, whilst methods ii)-iv) assume {\it a priori} that the PDF is log-normal, though for methods ii) and iii) only for obtaining the linear variance. Method iv) is the usual approach used to fit $b$ for a given Mach number, if one additionally assumes the relationship given by Eq.~(\ref{eq:bfac}) --- an assumption we do not need to make here since a range of Mach numbers are examined.

 While the results are discussed in more detail below, essentially we find that methods ii)-iv) all give similar results for $\sigma_{s}$, independent of numerical resolution, but that direct measurements of $\sigma_{\rho/\bar{\rho}}$ [method i)] are highly resolution-dependent.
 
%
 Volume-weighted variances were computed from the (mass weighted) SPH data by interpolating the density field to a grid. We found that this procedure gave better results than the direct calculation from the particles we have previously advocated \citepalias{pf10}, particularly at high Mach numbers ($\mathcal{M}\gtrsim10$) where assuming that the volume element $m/\rho$ is constant over the smoothing radius is an increasingly poor approximation. However, capturing the full resolution in the density field was found to require an adaptive rather than fixed mesh.
 
 All plots show volume-weighted quantities, time-averaged over 81 snapshots evenly spaced between $t/t_{d}=2$ and $t/t_{d}=10$, with error bars showing the (temporal) $1\sigma$ deviations from these values.
%

\begin{figure}
\begin{center}
 \includegraphics[width=\columnwidth]{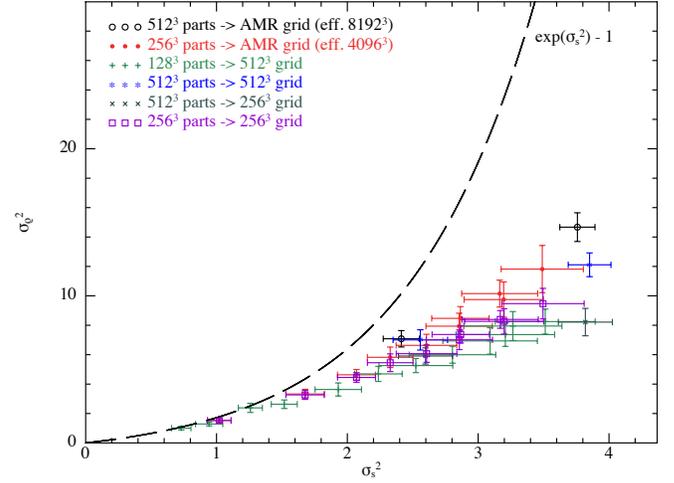}
 \caption{Relationship between the linear and logarithmic density variance as a function of both intrinsic SPH resolution (number of particles) and the grid size used to compute the variances (see legend). While the measurements of $\sigma_{s}^{2}$ are resolution independent, there is a strong dependence on both the SPH and grid resolution in the directly measured linear variance, $\sigma_{\rho/\bar{\rho}}^{2}$. Using an AMR grid to compute volume-weighted variances captures the full density field resolution in the SPH simulations, but even in the highest resolution calculations ($512^{3}$ particles), $\sigma_{\rho/\bar{\rho}}^{2}$ is severely underestimated compared to the expected exponential relationship (Eq.~\ref{eq:sigmas}, dashed line) for $\mathcal{M}\gtrsim5$.}
\label{fig:varsvarho}
\end{center}
\end{figure}

\begin{figure*}
\begin{center}
 \includegraphics[width=0.65\textwidth]{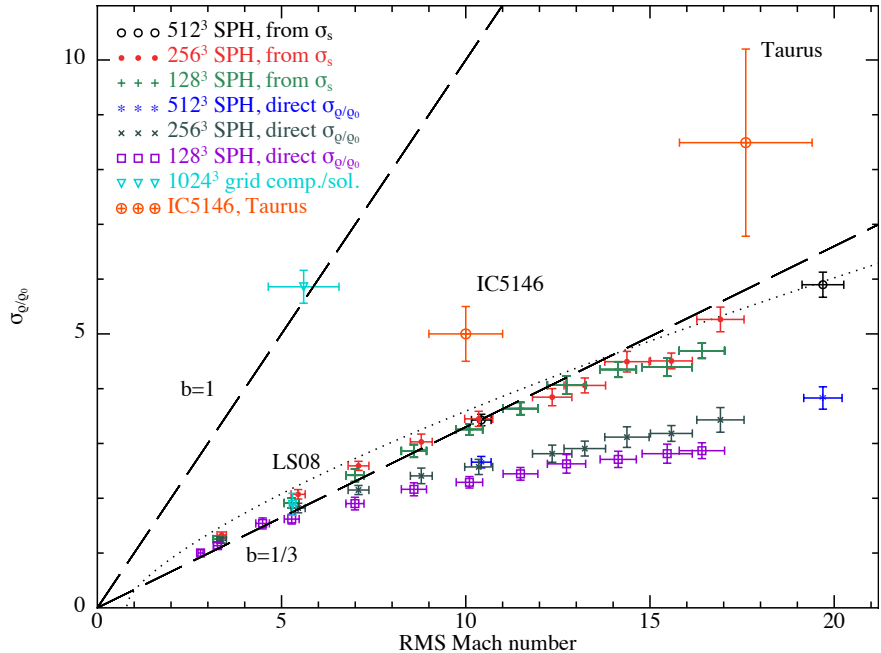}
 \caption{Directly measured or inferred (see legend) standard deviation of the linear density $\sigma_{\rho/\bar{\rho}}$ as a function of RMS Mach number from the solenoidally-driven supersonic turbulence calculations. For comparison, observational determinations by \citet{pjn97} and \citet{brunt10} in IC5146 and Taurus (respectively) are shown, together with the expected $b=1/3$ and $b=1$ linear relationships for solenoidal and compressive forcing (respectively) \citep{fks08,federrathetal10}, including the corresponding data points from \citet{federrathetal10} ($1024^{3}$ grid; cyan triangles). Direct measurements of $\sigma_{\rho/\bar{\rho}}$ are resolution-limited (see Fig.~\ref{fig:varsvarho}), although the values inferred by assuming Eq.~(\ref{eq:sigmas}) are \emph{upper} limits, whereas the observations are likely to be \emph{lower} limits. The discrepancy between solenoidally-driven simulations and observations indicates that some amount of gravity and/or compressive driving is necessary to explain the observational results.}
\label{fig:sigmarho}
\end{center}
\end{figure*}

\section{Density variance -- Mach number relation in supersonic, isothermal turbulence}  

\subsection{$\sigma_{s}$ as a function of $\mathcal{M}$}
 The direct measurements of the standard deviation in the log density, $\sigma_{s}$, are shown in Fig.~\ref{fig:sigmamach} from the results of calculations performed using $128^{3}$, $256^{3}$ and $512^{3}$ SPH particles, using methods ii) and iv) (see legend). Dashed lines show the standard relation (Eq.~\ref{eq:bfac}) with $b=1/3$ and $b=1/2$, whilst the dotted line shows the best fitting relationship found by \citetalias{ls08} (Eq.~\ref{eq:ls08}). Both the $b=1/3$ curve, expected for solenoidally-driven turbulence \citep{fks08,federrathetal10} and the \citetalias{ls08} fit show reasonable fits to the data, and indeed cannot be distinguished given the time variability present in the calculations. However, adopting $b=0.5$ is clearly not consistent with our solenoidally-driven results. The results are also consistent with our earlier findings \citepalias{pf10}, that showed a convergence in both grid and SPH methods towards $b\approx0.35-0.4$ at Mach 10.

 The measured value of $\sigma_{s}$ cannot be compared to observations, since only the 3D variance in the linear density can be observationally inferred (e.g. using the \citetalias{bfp10a} method). Thus it is necessary to either measure or infer the linear variance from simulations to make this comparison. 

\subsection{Direct measurement of $\sigma_{\rho/\bar{\rho}}$}
A direct measurement of the linear density variance is difficult even with high resolution simulations, demonstrated in Fig.~\ref{fig:varsvarho} that shows the directly measured $\sigma_{\rho/\bar{\rho}}^{2}$ as a function of $\sigma_{s}^{2}$. The dashed line shows the expected relationship for a log-normal distribution (Eq.~\ref{eq:sigmas}). The exponential relationship is resolved only at the lowest Mach numbers, $\mathcal{M}\lesssim 5$, corresponding to $\sigma_{s}^{2}\lesssim1.4$, while at higher Mach numbers $\sigma_{\rho/\bar{\rho}}$ is a strong function of resolution, most easily demonstrated by interpolating the highest resolution SPH calculations ($512^{3}$ particles) to fixed grids of decreasing resolution (see legend). This dependence is the reason why we eventually interpolated the SPH data onto an adaptive mesh, refined such that $\Delta x<h$ for all cells within the smoothing radius $2h$ of any given particle, in order to obtain a result for $\sigma_{\rho/\bar{\rho}}$ that captures the maximum resolution available in the SPH simulations, though we remain limited by the intrinsic resolution of the simulations. \citetalias{pf10} found that SPH simulations at Mach~10 resolved a maximum density at $128^{3}$ particles similar to that captured on a fixed grid at $512^{3}$ grid cells. This is consistent with the results here, where it is necessary to refine the grid to an effective $8192^{3}$ for the Mach~20 calculations employing $512^{3}$ SPH particles. It is also evident that fully resolving the strong fluctuations in the linear density at high Mach number is intractable with current computational resources. Our findings also suggest that the linear density variance is likely to be severely underestimated by limited observational resolution, so the results in Taurus and IC5146 are almost certainly lower limits \citep[see also][]{brunt10}.

\subsection{$\sigma_{\rho/\bar{\rho}}$ as a function of $\mathcal{M}$: Comparison to observations}
 The direct -- but resolution limited -- measurements for $\sigma_{\rho/\bar{\rho}}$, computed without assumptions from the AMR grid, are shown in Fig.~\ref{fig:sigmarho} (see legend) [i.e., method i), above]. A better approach is to use the fact that the measurements for $\sigma_{s}$ are resolution independent (Fig.~\ref{fig:sigmamach}), meaning that we can use the assumption of log-normality to infer the fully resolved value for $\sigma_{\rho/\bar{\rho}}$, i.e., using Eq.~(\ref{eq:sigmas}) (method ii). The standard deviations computed in this way are also shown, and as expected are consistent with a linear $\sigma_{\rho/\bar{\rho}}$--$\mathcal{M}$ relation with $b=1/3$ for solenoidal forcing \citep{fks08,federrathetal10}, and also with the \citetalias{ls08} best fit, the latter translated in terms of $\sigma_{\rho/\bar{\rho}}$ using Eq.~(\ref{eq:sigmas}).

  We are now in a position to compare with the observational results. It should first be noted that the assumption of the $\sigma_{s}$--$\sigma_{\rho/\bar{\rho}}$ relation for the simulations essentially gives us an \emph{upper} limit on $\sigma_{\rho/\bar{\rho}}$, whereas (see above) the finite resolution of the observations almost certainly gives a \emph{lower} limit on the variance. The results from \citet{pjn97} (in IC5146: $b=0.5\pm 0.05$ at Mach~10) and \citet{brunt10} (in Taurus, $\mathcal{M}_{\rm RMS}=17.6\pm1.8$ and $\sigma_{\rho/\bar{\rho}}=8.49^{+1.85}_{-1.71}$) are also plotted, both of which are consistent with $b=0.5$ but clearly inconsistent with the calculations of purely solenoidally-driven turbulence.
  
 The other extreme is given by the $b=1$ line in Fig.~\ref{fig:sigmarho}, corresponding to purely compressive forcing \citep{fks08,federrathetal10}. Data points from two $1024^3$ grid simulations of purely solenoidal and purely compressive forcing by \citet{fks08,federrathetal10} are also shown (cyan triangles). Clearly, purely solenoidal and purely compressive forcing seem inconsistent with the observations, while a mixture and/or the addition of gravity can fit the observations, best fit by a linear relation with $b\approx0.5$. The $\sigma_\rho$--$\mathcal{M}$ plane, however, needs to be populated with many more observational measures to draw more definite conclusions about, e.g., regional and evolutionary variations.
 \begin{figure}
\begin{center}
 \includegraphics[width=\columnwidth]{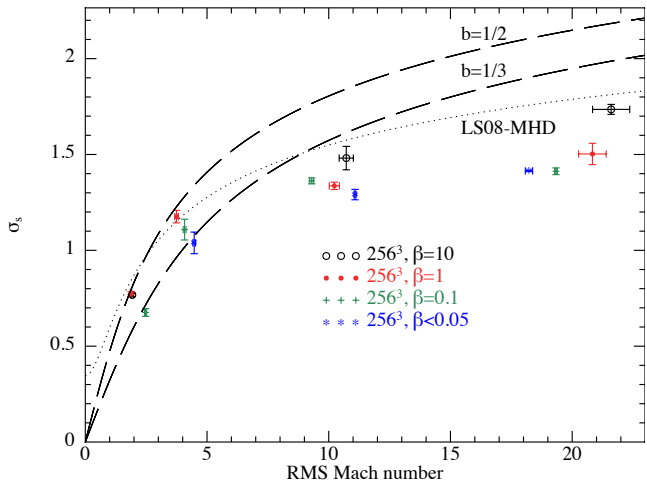}
 \caption{As in Fig.~\ref{fig:sigmamach} but for a series of $256^{3}$ grid-based MHD calculations with field strength characterised by the ratio of gas-to-magnetic pressure $\beta$ (see legend). There is a general decrease in the measured variance in the MHD simulations at high Mach number, though no clear trend with magnetic field strength. The best-fitting relationship found by \citetalias{ls08} for strong field MHD calculations (dotted line) is consistent with our $\beta=1$ results in a similar parameter range ($\mathcal{M} \lesssim 6$), but too steep at higher $\mathcal{M}$. The $\beta<0.05$ points refer to calculations employing $\beta = 0.05$, $0.01$ and $0.02$ at Mach $4$, $10$ and $20$ respectively.}
\label{fig:MHD}
\end{center}
\end{figure}

\section{Density variance -- Mach number relation in MHD turbulence}
 Since molecular clouds are observed to be magnetised, we have additionally computed a series of Magnetohydrodynamics (MHD) calculations, driven identically to their hydrodynamic counterparts but with an initially uniform magnetic field threading the box. These have been computed on a fixed grid using the \textsc{flash} code, as described in \citet{federrathetal10}, \citetalias{pf10}, and \citet{bfp10a,bfp10b} (note that the driving routines are implemented identically in both the SPH and grid code). The magnetic field strength in the MHD calculations is characterised by the ratio of gas-to-magnetic pressure $\beta=P/ P_{mag}$ in the initial conditions, where $P_{mag}=\frac12 B^{2}/\mu_{0}$. As previously we have also examined the effect of resolution, with a finding similar to the hydrodynamic case -- namely that direct measurements of $\sigma_{\rho/\bar{\rho}}$ are strongly resolution affected (underestimated) -- even at modest Mach numbers, similar to the results shown in Fig.~\ref{fig:varsvarho}, but that measurements of $\sigma_{s}$ are resolution independent (at $\gtrsim 256^{3}$ grid cells).

 Fig.~\ref{fig:MHD} shows the results, similar to Fig.~\ref{fig:sigmamach} for the hydrodynamic case and over-plotted with the hydrodynamic $b=1/3$ and $b=1/2$ relationships (dashed lines) as well as the best-fitting MHD relationship found by \citetalias{ls08}. The most obvious difference with MHD is that the density variances are significantly lower than their hydrodynamic counterparts at high (sonic) Mach number $\mathcal{M}\gtrsim10$ -- though conversely marginally higher at lower Mach numbers. On the whole increasing the field strength seems to decrease the mean $\sigma_{s}$ slightly. Whilst a complete MHD study is beyond the scope of this paper, the results clearly illustrate a shallower relationship than the hydrodynamic $b=1/3$ curve at high Mach number, with $\sigma_{s}$ only weakly dependent on $\mathcal{M}$ in this regime (for $\beta\lesssim1$). Thus, if anything, adding magnetic fields decreases the variance in the density field, worsening the discrepancy with observations.

\section{Conclusions}
 We have measured the relationship between the density variance and the mean square Mach number from a series of simulations of supersonic, isothermal, solenoidally-driven turbulence over a wide range of Mach numbers ($1\lesssim\mathcal{M}\lesssim20$). We find that the standard relationship given by Eq.~(\ref{eq:bfac}) with $b=1/3$ provides a good fit to the data over this range, consistent with the heuristic theory of \citet{federrathetal10} for solenoidal driving and with similar measurements by \citet{ls08} at lower Mach numbers. While it is difficult to measure the variance in linear density directly from simulations with finite resolution, the inferred relationship (Eq.~\ref{eq:bfaclin}, with $b=1/3$) appears inconsistent with observational determinations ($b\approx 0.5$) in Taurus and IC5146, suggesting that additional physics such as gravity is important in these clouds and/or that some form of compressive driving is relevant. This is consistent with the findings of \citet{kainulainenetal09} and \citet{lombardietal10} for Taurus, where self-gravity is invoked to explain the deviation from log-normality in the high density tail of the (column) density PDF. Magnetic fields do not help to explain the discrepancy.

\section*{Acknowledgments}
 We acknowledge computational resources from the Monash Sun Grid (with thanks to Philip Chan) and the Leibniz Rechenzentrum Garching. Plots utilised \textsc{splash} \citep{splashpaper} with the new giza backend by James Wetter. CF is supported by Baden-W\"{u}rttemberg Stiftung grant~P-LS-SPII/18.

\label{lastpage}
\enddocument